\documentstyle[emulateapj]{article}
\newcommand{\kms}{km s$^{-1}$}
\newcommand{\zabs}{$z_{\rm abs}$}
\newcommand{\lya}{Ly$\alpha$\ }
\slugcomment{Accepted for publication in the {\it Ap.J. Letters}}
\lefthead{TRIPP ET AL.}
\righthead{BARYONIC CONTENT OF O VI ABSORBERS}
\begin{document}
\title{Intervening \ion{O}{6} Quasar Absorption Systems at Low Redshift: \nl
A Significant Baryon Reservoir\altaffilmark{1}}

\altaffiltext{1}{Based on observations with the NASA/ESA {\it Hubble 
Space Telescope}, obtained at the Space Telescope Science Institute, which 
is operated by the Association of Universities for Research in Astronomy, 
Inc., under NASA contract NAS 5-2555. The data were reduced with the STIS 
Team software, and the research was supported by NASA through grants 
GO-08165.01-97A and GO-08165.02-97A from STScI.}

\author{Todd M. Tripp,\altaffilmark{2} Blair D. Savage,\altaffilmark{3} 
and Edward B. Jenkins\altaffilmark{2}}

\altaffiltext{2}{Princeton University Observatory,
Peyton Hall, Princeton, NJ 08544,
Electronic mail: tripp@astro.princeton.edu, ebj@astro.princeton.edu}

\altaffiltext{3}{Department of Astronomy, University of Wisconsin-Madison,
Madison, WI 53706
Electronic mail: savage@astro.wisc.edu}

\begin{abstract}
Far-UV echelle spectroscopy of the radio-quiet QSO H1821+643 ($z_{\rm em}$ = 
0.297), obtained with the Space Telescope Imaging Spectrograph (STIS) at 
$\sim$7 \kms\ resolution, reveals 4 definite \ion{O}{6} absorption line systems
and one probable \ion{O}{6} absorber at 0.15 $<$ \zabs\ $<$ 0.27.  The four 
definite \ion{O}{6} absorbers are located near galaxies and are highly 
displaced from the quasar in redshift; these are likely intervening systems 
unrelated to the background QSO. In the case of the strong \ion{O}{6} system 
at \zabs\ = 0.22497, multiple components are detected in \ion{Si}{3} and 
\ion{O}{6} as well as \ion{H}{1} Lyman series lines, and the differing 
component velocity centroids and $b-$values firmly establish that this is a 
multiphase absorption system. A weak \ion{O}{6} absorber is detected at \zabs\ 
= 0.22637, i.e., offset by $\sim$340 \kms\ from the \zabs\ = 0.22497 system. 
\lya\ absorption is detected at \zabs\ = 0.22613, but no \lya\ absorption is 
significantly detected at 0.22637.  Other weak \ion{O}{6} absorbers at \zabs\ = 
0.24531 and 0.26659 and the probable \ion{O}{6} system at 0.21326 have widely 
diverse \ion{O}{6}/\ion{H}{1} column density ratios with 
$N$(\ion{O}{6})/$N$(\ion{H}{1}) ranging from $\leq 0.14\pm 0.03$ to 
$5.2\pm 1.2$.  The number density of \ion{O}{6} absorbers with rest 
equivalent width $>$ 30 m\AA\ in the H1821+643 spectrum is remarkably high, 
$dN/dz \sim$ 48, with a high (90\%) confidence that it is greater than 17. We 
conservatively estimate that the cosmological mass density of the 
\ion{O}{6} systems $\Omega _{b}({\rm O \ VI}) \gtrsim 0.0008 h_{75} ^{-1}$. 
With an assumed metallicity of 1/10 solar and a conservative assumption that 
the fraction of oxygen in the \ion{O}{6} ionization stage is 0.2, we obtain 
$\Omega _{b}({\rm O \ VI}) \gtrsim 0.004 h_{75} ^{-1}$. This is comparable to 
the combined cosmological mass density of stars and cool gas in galaxies and 
X-ray emitting gas in galaxy clusters at low redshift.
\end{abstract}


\section{Introduction}

The resonance line doublet of Li-like \ion{O}{6} is a sensitive probe of hot 
collisionally ionized or warm very low density photoionized gas in the 
intergalactic medium and galaxy halos.  The \ion{O}{6} $\lambda 
\lambda$1031.92, 1037.62 doublet has been detected in 
absorption toward QSOs over a wide range of
redshifts (see \S 1 in Tripp \& Savage\markcite{ts2000} 2000). The lowest
redshift \ion{O}{6} absorbers are particularly interesting because the 
redshifts of galaxies near the QSO sight lines can be measured, and the 
relationship between the \ion{O}{6} absorber properties and environment can
be studied.  Furthermore, cosmological simulations predict that a substantial
fraction of the baryons in the universe are in a shock-heated phase at 
$10^{5} - 10^{7}$ K at low $z$ (e.g., Cen \& Ostriker\markcite{co99} 1999; 
Dav\'{e} et al.\markcite{dave99} 1999), and preliminary results indicate that 
low-$z$ \ion{O}{6} systems may indeed be an 
important baryon reservoir (Tripp \& Savage\markcite{ts2000} 2000).
In a previous paper, Savage, Tripp, \& Lu\markcite{stl98} (1998) studied
an intervening \ion{O}{6} absorber associated with two galaxies at $z \approx$ 
0.225 in the spectrum of the radio-quiet QSO H1821+643 using a combination of 
low resolution {\it Hubble Space Telescope (HST)} spectra with broad wavelength 
coverage and a high resolution {\it HST} spectrum with very limited wavelength
coverage. We have re-observed this QSO with an echelle mode of the Space 
Telescope Imaging Spectrograph (STIS) on {\it HST}, which provides a resolution
of $\sim$7 \kms\ (FWHM) with broad wavelength coverage. In this paper we 
present in \S 2 and \S 3 new results on one probable and four definite 
\ion{O}{6} absorption line systems in the STIS H1821+643 spectrum. In \S 4 we 
discuss the implications of the high rate of occurance of \ion{O}{6} absorbers 
at low redshift. The direct information we obtain about the highly ionized 
state of the gas from the presence of \ion{O}{6} allows us to estimate the 
baryonic content of these systems. We conclude that \ion{O}{6} systems are 
likely to harbor an important fraction of the baryons at the present epoch.

\section{Observations and Absorption Line Measurements}

H1821+643 was observed with STIS for 25466 seconds on 1999 June 25 with the 
medium resolution FUV echelle mode (E140M) and the 0.2$\times$0.06'' 
slit.\footnote{{\it HST} archive ID numbers O5E703010--O5E7030E0.} This STIS 
mode provides a resolution of $R \ = \ \lambda /\Delta \lambda \approx$ 46000
or FWHM $\approx$ 7 \kms\ (Kimble et al.\markcite{kim98} 1998). 
The data were reduced as described by Tripp \& Savage\markcite{ts99} 
(2000) including the scattered light correction developed by the STIS 
Instrument Definition Team. The spectrum extends from $\sim$1150 to 1710 \AA\ 
with four small gaps between orders at $\lambda >$ 1630 \AA . Throughout 
this paper wavelengths and redshifts are heliocentric.

\begin{figure}
\plotone{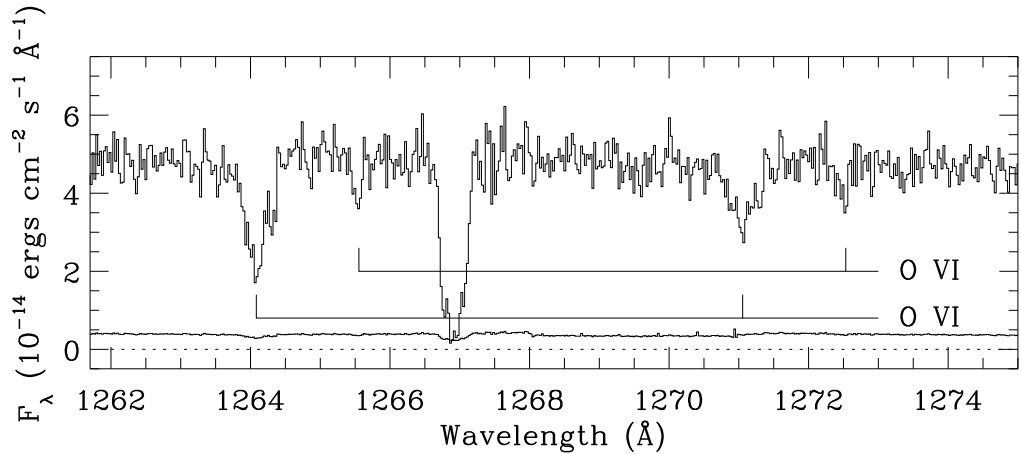}
%\plotfiddle{smp2.ps}{4.0in}{0}{55}{55}{-200}{-70}
\caption[]{Portion of the STIS E140M spectrum of H1821+643 showing the 
strong \ion{O}{6} absorption lines at \zabs\ = 0.22497 and the weaker \ion{O}{6}
absorber at \zabs\ = 0.22637. The calibrated flux is plotted vs. observed 
heliocentric wavelength, and the solid line near zero is the $1\sigma$ flux 
uncertainty. The line at 1266.9 \AA\ is an unrelated \ion{C}{3} line from the 
absorption system at \zabs\ = 0.2967. In this figure, the
spectrum has been binned 2 pixels $\rightarrow$ 1 pixel for display purposes
only (all measurements in the text were made using the unbinned full resolution
data).\label{samplespec}}
\end{figure}

We first searched the spectrum for \ion{O}{6} absorbers by checking for lines 
with the velocity separation and relative line strengths expected for the 
doublet. This identified the four definite \ion{O}{6} systems. We then 
searched for \ion{O}{6} lines associated with known \lya absorbers, and this
revealed the probable system (see below).  A selected sample of the spectrum 
is shown in Figure ~\ref{samplespec}. This portion of the spectrum shows the 
\ion{O}{6} doublet at \zabs\ = 0.22497 as well as a much weaker \ion{O}{6} 
doublet at \zabs\ = 0.22637. Both of these \ion{O}{6} absorbers are discussed 
in \S 3.1. In addition to the \ion{O}{6} systems in Figure ~\ref{samplespec}, 
the STIS echelle spectrum shows new \ion{O}{6} absorbers at \zabs\ = 0.24531 
and 0.26659 which have small equivalent widths; these systems are briefly 
discussed in \S 3.2 along with a possible \ion{O}{6} system at \zabs\ = 
0.21326.  The \zabs\ = 0.21326 system is a strong \ion{H}{1} 
Ly$\alpha$/Ly$\beta$ absorber with a $> 4\sigma$ line detected at the 
expected wavelength of \ion{O}{6} $\lambda$1031.93. However, the corresponding 
\ion{O}{6} $\lambda$1037.62 line is blended with Milky Way \ion{S}{2} 
$\lambda$1259.52 absorption due to two high velocity clouds, so we consider 
this a probable but not definite \ion{O}{6} detection. The component structure 
establishes that this blend is mostly due to Milky Way \ion{S}{2}, but it is 
possible that an \ion{O}{6} $\lambda$1037.62 line of the right strength is 
present as well. In principle this could be proved by comparing the \ion{S}{2} 
$\lambda$1259.52 line strengths to the \ion{S}{2} $\lambda \lambda$1250.58, 
1253.81 line strengths. However, this does not yield a clear result at the 
current S/N level due to ambiguity of the continuum placement near 1259 \AA .  

Restframe equivalent widths ($W_{\rm r}$) of absorption lines detected in the
\ion{O}{6} systems, measured using the software of Sembach \& 
Savage\markcite{ss92} (1992), are listed in Table ~\ref{lineprop}.  Note that 
the quoted errors in equivalent width include contributions from uncertainties 
in the height and curvature of the continuum as well as a 2\% uncertainty 
in the flux zero point. 
Integrated apparent column densities (Savage \& Sembach\markcite{ss91} 1991) 
are also found in Table ~\ref{lineprop} with error bars including 
contributions from continuum and zero point uncertainties. 
To measure line widths, we used the Voigt profile fitting software of 
Fitzpatrick \& Spitzer\markcite{fitz97} (1997) with the line spread functions 
from the Cycle 9 STIS Handbook.

\section{Absorber Properties}

Four of the five absorption systems in Table ~\ref{lineprop} are within a
projected distance of 1 $h_{75}^{-1}$ Mpc or less of at least one galaxy with 
$\mid \Delta v\mid \ = \mid c(z_{\rm gal} - z_{\rm abs})/(1 + z_{\rm mean}) 
\mid \ \leq$ 300 \kms , and some of them are close to multiple 
galaxies (see Table 1 in Tripp et al.\markcite{tls98} 1998). These absorbers 
are also displaced from the QSO redshift by 7100 \kms\ (\zabs\ = 0.26659) to 
17100 \kms\ (\zabs\ = 0.22497). Finally, the \ion{O}{6} profiles are 
relatively narrow. Therefore these are probably intervening 
systems that trace the large-scale gaseous environment in galaxy envelopes 
and the IGM rather than ``intrinsic'' absorbers (Hamann \& 
Ferland\markcite{hf99} 1999).

\subsection{\ion{O}{6} Absorbers at z = 0.22497 and 0.22637}

\begin{figure}
%\plotone{stack_onep22497.eps}
\plotfiddle{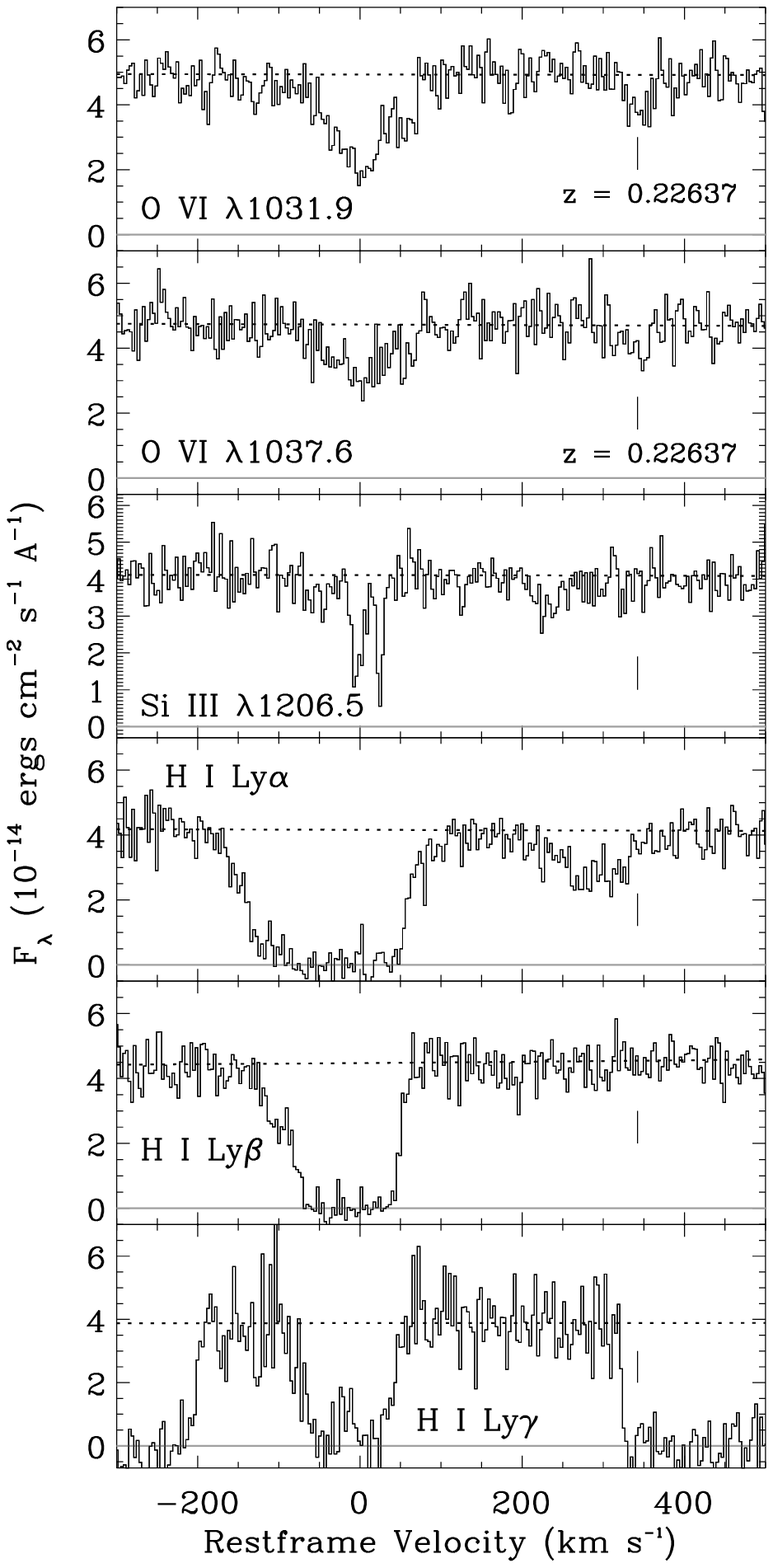}{6.0in}{0}{72}{72}{-200}{-150}
\caption[]{Profiles of absorption lines detected in the \ion{O}{6} system 
at \zabs\ = 0.22497, plotted versus restframe velocity where $v$ = 0 at
\zabs\ = 0.22497. The tick mark at $v \approx$ 340 \kms\ indicates the velocity 
of the \zabs\ = 0.22637 absorber in the \zabs\ = 0.22497 restframe. The dotted 
lines show the continua adopted for absorption line measurements, and the grey 
lines show the flux zero levels. The data in this figure are shown at full 
resolution (no binning has been applied). Note that the strong lines at $v 
<$ $-$200 and $>$ 330 \kms\ in the \ion{H}{1} Ly$\gamma$ panel are 
due to the ISM \ion{Si}{2} 1190.42 and 1193.29 \AA\ lines, respectively.
\label{stack22497}}
\end{figure}

Since the \ion{O}{6} doublets at \zabs\ = 0.22497 and 0.22637 shown in Figure
~\ref{samplespec} are separated by only $\sim$340 \kms , they are probably 
related and we discuss them together. Two emission line galaxies are known at 
heliocentric redshifts of 0.22560 and 0.22650 at projected distances of 105 
and 388 $h_{75}^{-1}$ kpc from the sight line\footnote{In this paper, the 
cosmological parameters are set to $H_{0} = 75 h_{75}$ \kms\ Mpc$^{-1}$ and 
$q_{0}$ = 0.0.} (Tripp et al.\markcite{tls98} 1998).  In addition to 
the \ion{O}{6} doublet, the STIS spectrum shows strong absorption lines due to 
\ion{H}{1} \lya , Ly$\beta$, Ly$\gamma$, \ion{Si}{3} $\lambda$1206.5, and 
possibly \ion{C}{3} $\lambda$977.02 at \zabs\ = 0.22497; the absorption 
profiles of most of these species are plotted on a velocity scale in 
Figure ~\ref{stack22497}. The lines of \ion{N}{5} and \ion{Si}{4} are not 
detected at greater than 3$\sigma$ significance, and upper limits on their 
equivalent widths and column densities are listed in Table ~\ref{lineprop} 
along with upper limits on \ion{C}{2} and \ion{Si}{2}. 

Figure ~\ref{stack22497} provides several indications that these 
systems are multiphase absorbers. Several components are readily apparent in 
most of the \zabs\ = 0.22497 profiles including the \ion{O}{6} lines (see 
also Figure ~\ref{samplespec}). Fitting of the \ion{Si}{3} profile yields $b$ =
7.7$^{+2.9}_{-2.1}$ and 1.6$^{+2.1}_{-0.9}$ \kms\ for the two well-detected
components at $v$ = --3 and +25 \kms , respectively. However, the component 
velocity centroids and $b$-values are not compatible with a homogeneous 
mixture of \ion{O}{6} and \ion{Si}{3}.  For example, the \ion{Si}{3} profile 
shows a prominent narrow component at $v \approx$ 25 \kms , and there is no 
obviously corresponding component in the \ion{O}{6} profiles. While thermal 
Doppler broadening can make the \ion{O}{6} profiles broader than those of 
\ion{Si}{3}, at most the increase will be a factor of $\sqrt{28/16}$, and this
is inadequate to produce the breadth of the observed \ion{O}{6} lines. 
Thus we are compelled to consider a mixture of phases, some of which show 
up in \ion{Si}{3}, while others are prominent in \ion{O}{6}.

In the case of the \ion{O}{6} at \zabs\ = 0.22637, which is also visible in 
Figure ~\ref{stack22497}, the multiphase nature is suggested by an offset of 
60 \kms\ between the \ion{H}{1} \lya\ and \ion{O}{6} velocity centroids.  
Also we note that no \ion{H}{1} absorption is significantly detected at the 
velocity of the \ion{O}{6} suggesting that the hydrogen is thoroughly ionized 
in the \ion{O}{6} gas. This \ion{O}{6} absorber may be analogous to the highly 
ionized high velocity clouds seen near the Milky Way which show strong high ion 
absorption with very weak or absent low ion absorption (Sembach et 
al.\markcite{sem99} 1999). 

\subsection{Other Weak \ion{O}{6} Systems}

\begin{figure}
%\plotone{stackweak.eps}
\plotfiddle{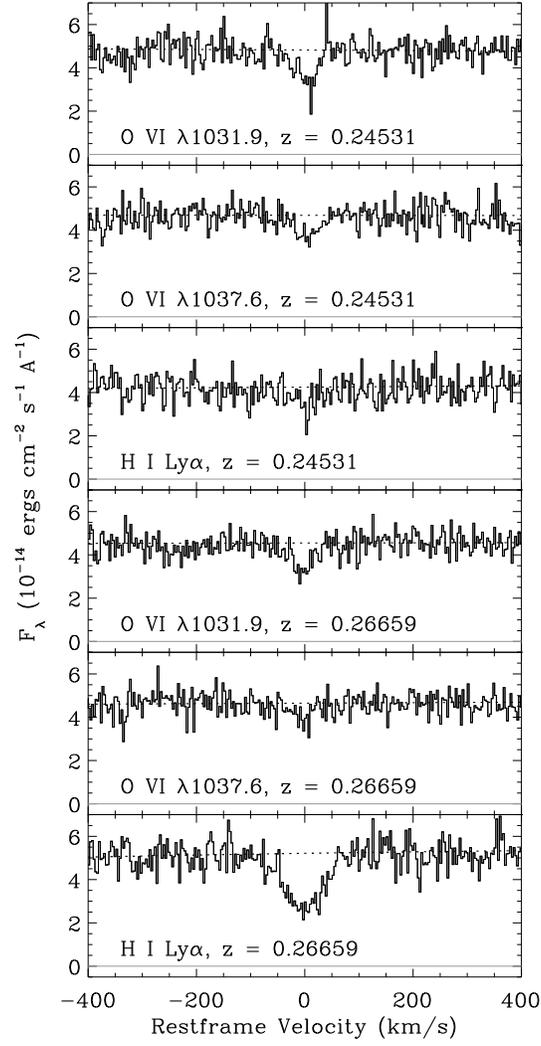}{6.0in}{0}{72}{72}{-200}{-150}
\caption[]{Absorption profiles of lines detected in the \ion{O}{6} absorbers 
at \zabs\ = 0.24531 (upper three panels) and at \zabs\ = 0.26659 (lower 
three panels), plotted versus restframe velocity. \label{weak}}
\end{figure}

The two new \ion{O}{6} systems at \zabs\ = 0.24531 and \zabs\ = 0.26659 are 
plotted in Figure ~\ref{weak}. A striking feature of these weak \ion{O}{6} 
absorbers (and the candidate \ion{O}{6} at \zabs\ = 0.21326) is that while 
their \ion{O}{6} column densities are comparable, the strengths of their 
corresponding \ion{H}{1} absorption lines are significantly different (see 
Table ~\ref{lineprop} and Figure ~\ref{weak}).  For example, 
$N$(\ion{O}{6})/$N$(\ion{H}{1}) = 5.2$\pm 1.2$ in the \zabs\ = 0.24531 system 
while $N$(\ion{O}{6})/$N$(\ion{H}{1}) = 1.2$\pm 0.2$ in the \zabs\ = 0.26659 
absorber.  The contrast is even more dramatic with the \zabs\ = 0.21326 
absorber, which has $N$(\ion{O}{6})/$N$(\ion{H}{1}) $\leq 0.14\pm 0.03$.
For reference, in collisional ionization equilibrium (Sutherland \& 
Dopita\markcite{sd93} 1993), gas with solar metallicity at the peak 
\ion{O}{6} ionization temperature should have 
$N$(\ion{O}{6})/$N$(\ion{H}{1}) $\sim$ 100. 
The large variablility of the observed \ion{O}{6}/\ion{H}{1} ratio could 
indicate that the metallicity of the \ion{O}{6} absorbers varies substantially,
or this could be due to differences in the physical conditions 
and ionization of the gas. If, for example, these are multiphase absorbers 
with the \ion{H}{1} lines arising in a cool phase which is embedded in a hot 
phase which produces the \ion{O}{6} absorption (e.g., Mo \& 
Miralda-Escud\'{e}\markcite{mo} 1996), then the wide variations in the 
\ion{O}{6}/\ion{H}{1} ratio could simply be due to interception of fewer cool 
phase clouds in one absorber compared to another. 

A full analysis of the range of physical conditions of these absorbers will be
presented in a later paper. However, it is interesting to note that the 
\ion{H}{1} \lya profile of the \zabs\ = 0.26659 system is rather broad and 
relatively smooth (see the bottom panel of Figure ~\ref{weak}), which may 
indicate that this absorber is collisionally ionized and hot. However, fitting 
a single component to the \zabs\ = 0.26659 \lya profile yields 
$b = 44.6^{+7.3}_{-6.3}$, which implies that $T \lesssim 1.2 \times 10^{5}$ K. 
At this temperature the \ion{O}{6} ionization fraction is rather small in 
collisional ionization equilibrium (Sutherland \& Dopita\markcite{sd93} 1993), 
and an unreasonably high metallicity is required to produce the observed 
$N$(\ion{O}{6}) and $N$(\ion{H}{1}) in the same gas. This may be another 
indication that these are multiphase absorbers or that the gas is not in 
ionization equilibrium.

\section{Number Density and Cosmological Mass Density}

The new STIS data in this paper provide an opportunity to evaluate
the number density of low-$z$ \ion{O}{6} absorbers per unit redshift 
($dN/dz$) and a lower limit on their cosmological mass density. If we neglect 
continuum placement uncertainty and other systematic error sources, the STIS 
E140M spectrum of H1821+643 is formally adequate for $\geq 4\sigma$ detection 
of narrow lines with $W_{\rm r} \geq$ 30 m\AA\ at $\lambda _{\rm obs} 
\gtrsim$ 1188 \AA\ (\zabs\ $\gtrsim$ 0.151 for \ion{O}{6} $\lambda$1031.93). 
However, the continuum placement ambiguity substantially increases the 
uncertainty in $W_{\rm r}$ for weak lines. Moreover, 
broader resolved lines spread over more pixels have higher limiting equivalent 
widths (limiting $W \propto \sqrt{\rm no. pixels}$), so broad weak lines may 
not be detected at the $4 \sigma$ level. Consequently, the $dN/dz$ derived 
below should be treated as a lower limit. We require detection of both lines 
of the \ion{O}{6} doublet with $W_{\rm r} \geq$ 30 m\AA , and we exclude one 
absorber\footnote{We exclude the associated O VI absorber at \zabs\ = 
0.2967. This system is not listed in Table ~\ref{lineprop} but is discussed in 
detail in Savage et al.\markcite{stl98} (1998) and Oegerle et 
al.\markcite{bill} (2000).} within $\mid \Delta v \mid \ \leq$ 5000 \kms\ of 
$z_{\rm em}$ to avoid contamination of the sample with intrinsic absorbers.  
This results in a sample of three \ion{O}{6} systems\footnote{The three 
systems include those at \zabs\ = 0.22497, 0.24531, and 0.26659. We exclude 
the probable system at \zabs\ = 0.21326, and the \zabs\ = 0.22637 system 
falls below the equivalent width threshold.} over a redshift path of 
$\Delta z$ = 0.063 (after correction for a loss of $\Delta z$ = 0.061 for 
spectral regions in which either of the \ion{O}{6} lines is blocked by ISM or 
extragalactic lines from other redshifts). Therefore the most probable $dN/dz 
\sim$ 48 for $W_{\rm r} \geq$ 30 m\AA\ and 0.15 $\leq$ \zabs\ $\leq$ 0.27, and 
we conservatively conclude that $dN/dz \geq$ 17 at the 90\% confidence level 
(following the Gehrels\markcite{geh} 1986 treatment for small sample 
statistics). This is a remarkably high number density. It is important to
emphasize that the sample is extremely small and, since very little is known
about {\it weak} \ion{O}{6} lines at low redshift, it remains possible that 
$dN/dz$ is unusually high toward H1821+643 for some reason. However, there is 
supporting evidence that $dN/dz$ is generally high: (1) a similar $dN/dz$ is 
derived from STIS echelle spectroscopy of PG0953+415 (Tripp \& 
Savage\markcite{ts2000} 2000), and (2) one or two additional intervening 
\ion{O}{6} absorbers are evident in the H1821+643 spectrum which did not 
satisfy the selection criteria to be included in the sample.  More observations
are needed to build the sample of weak \ion{O}{6} lines at low $z$.

For comparison, low to moderate redshift \ion{Mg}{2} absorbers with 
$W_{\rm r} \geq$ 20 m\AA\ have $dN/dz = 2.65 \pm 0.15$ (Churchill et 
al.\markcite{crcv99} 1999; see also Tripp, Lu, \& Savage\markcite{tls97} 
1997). The {\it stronger} \ion{O}{6} absorbers are less common; 
Burles \& Tytler\markcite{bt96} (1996) report $dN/dz = 1.0\pm 0.6$ for
\ion{O}{6} systems with $W_{\rm r} \geq$ 210 m\AA\ at $<z_{\rm abs}>$ = 0.9.
Evidently, the $dN/dz$ of the weak \ion{O}{6} lines is substantially 
larger than $dN/dz$ of other known classes of low $z$ metal absorbers and is 
more comparable to that of low $z$ weak \lya absorbers, which have $dN/dz \sim$ 
100 for $W_{\rm r} \geq$ 50 m\AA\ (Tripp et al.\markcite{tls98} 1998; 
Penton et al.\markcite{pss} 2000).

Following analogous calculations (e.g., Storrie-Lombardi et al.\markcite{sl96} 
1996; Burles \& Tytler\markcite{bt96} 1996),\footnote{Note that while 
Burles \& Tytler\markcite{bt96} (1996) calculated the cosmological mass 
density of the oxygen ions in O VI absorbers (which is quite small), 
they did not apply an ionization and metallicity correction to estimate the
total baryonic content of the O VI systems. Instead, they used this 
method to place a lower limit on the mean metallicity of the O VI 
systems.} the mean cosmological mass density in the \ion{O}{6} absorbers, in 
units of the current critical density $\rho _{c}$, can be estimated using
\begin{equation}
\Omega _{b}({\rm O \ VI}) = \frac{\mu m_{\rm H} H_{0}}{\rho _{c} c
f({\rm O \ VI})} \left( \frac{\rm O}{\rm H} \right)^{-1}_{\rm O \ VI}
\frac{\sum_{i} N_{i}({\rm O \ VI})}{\Delta X}
\end{equation}
where $\mu$ is the mean atomic weight (taken to be 1.3), $f$(\ion{O}{6})
is a representative \ion{O}{6} ionization fraction, (O/H)$_{\rm O \ VI}$ 
is the assumed mean oxygen 
abundance by number in the \ion{O}{6} absorbers, $\sum_{i} N_{i}$(\ion{O}{6}) 
is the total \ion{O}{6} column density from the $i$ absorbers, and $\Delta X$ 
is the absorption distance interval (Bahcall \& Peebles\markcite{bah69} 1969), 
corrected for blocked spectral regions. With the sample defined above, we have 
$\Omega _{b}({\rm O \ VI}) = 8.0 \times 10^{-5} f({\rm O \ VI})^{-1} 
10^{-[{\rm O/H}]} h_{75} ^{-1} $ where [O/H] = log (O/H) - log (O/H)$_{\odot}$.
To set a conservative lower limit on $\Omega _{b}$(\ion{O}{6}), we 
assume [O/H] = $-$0.3 and $f$(\ion{O}{6}) = 0.2 (which is
close to the maximum value in photo- or collisional ionization, see 
Tripp \& Savage\markcite{ts2000} 2000), which yields 
$\Omega _{b}$(\ion{O}{6}) $\geq 0.0008 h_{75} ^{-1}$. If we set the mean 
metallicity to a more realistic value such as [O/H] = $-$1, 
$\Omega _{b}$(\ion{O}{6}) increases to $\geq 0.004 h_{75} ^{-1}$. Similar 
lower limits on $\Omega _{b}$(\ion{O}{6}) have been derived by Tripp \& 
Savage\markcite{ts2000} (2000) using a slightly less sensitive sample based on 
STIS echelle spectroscopy of PG0953+415 and earlier Goddard High Resolution 
Spectrograph observations of H1821+643. The lower limit assuming (O/H) = 1/10 
solar is comparable to the combined cosmological mass density of stars, cool 
neutral gas, and X-ray emitting cluster gas at low redshift, $\Omega _{*} + 
\Omega _{\rm H~I \ 21cm} + \Omega _{\rm H_{2}} + \Omega _{\rm X-ray} \approx$ 
0.006 (Fukugita, Hogan, \& Peebles\markcite{fhp98} 1998). Though still 
uncertain due to the small sample,\footnote{For a discussion of the impact of 
small number statistics on the $\Omega _{b}$(O VI) estimates, see 
Tripp \& Savage\markcite{ts2000} (2000).} small redshift path probed, and 
uncertain (O/H)$_{\rm O \ VI}$, these preliminary lower limits
on $\Omega _{b}$(\ion{O}{6}) suggest that \ion{O}{6} absorbers contain an 
important fraction of the baryons in the low redshift universe. 

\acknowledgements

We thank Ken Sembach and Ed Fitzpatrick for sharing their software 
for the measurement of column densities and $b$-values.

{\footnotesize
\begin{deluxetable}{llccc}
\tablewidth{0pc}
\tablecaption{Equivalent Widths and Integrated Column Densities\label{lineprop}}
\tablehead{$\lambda _{\rm obs}$\tablenotemark{a} & Species & $\lambda _{0}$\tablenotemark{b} & $W_{\rm
r}$\tablenotemark{c} & $N_{\rm a}$\tablenotemark{d} \nl
 \ & \ & (\AA ) & (m\AA ) & ($10^{13}$ cm$^{-2}$) }
\startdata 
\multicolumn{5}{c}{\bf $\bf z_{\rm abs}$ = 0.21326} \nl
1474.93 & H I  & 1215.67 & 471$\pm$14 & $>19.4$\tablenotemark{e} \nl
1244.47 & H I  & 1025.72 & 134$\pm$9 & $24.7\pm 2.0$ \nl
1252.07 & O VI\tablenotemark{f} & 1031.93 & 38$\pm$9 & $3.55\pm 0.81$ \nl
\hline
\multicolumn{5}{c}{\bf $\bf z_{\rm abs}$ = 0.22497} \nl
1488.97 & H I  & 1215.67 & 852$\pm$22 & $>34.4$\tablenotemark{e} \nl
1256.41 & H I  & 1025.72 & 503$\pm$14 & $>155$\tablenotemark{e} \nl
1191.27 & H I  & 972.54  & 340$\pm$17 & $>266$\tablenotemark{e} \nl
1264.09 & O VI  & 1031.93 & 185$\pm$9 & $19.9\pm 1.2$ \nl
1271.05 & O VI  & 1037.62 & 110$\pm$10 & $21.0\pm 2.0$ \nl
\nodata & Si II & 1260.42 & $<$47\tablenotemark{g} & $<$0.31\tablenotemark{g} \nl
1477.93 & Si III & 1206.50 & 108$\pm$9 & $0.73\pm 0.07$ \nl
\nodata & Si IV & 1393.76 & 48$\pm$23\tablenotemark{h} & $<$1.3\tablenotemark{h} \nl
\nodata & C II\tablenotemark{i}  & 1036.34 & $<$40\tablenotemark{g} & $<$3.6 \nl
1196.79 & C III?\tablenotemark{j} & 977.02 & 319$\pm$16 & $\geq 8.9$\tablenotemark{d} \nl
\nodata & N V & 1238.82 & $<$52\tablenotemark{g} & $<$2.5\tablenotemark{g} \nl
\hline
\multicolumn{5}{c}{\bf $\bf z_{\rm abs}$ = 0.22637} \nl
1490.57 & H I & 1215.67 & 168$\pm$15 & $3.7\pm 0.3$ \nl
1265.53 & O VI & 1031.93 & 25$\pm$5   & $2.4\pm 0.5$ \nl
1272.49 & O VI & 1037.62 & 21$\pm$5   & $3.7\pm 1.0$ \nl
\hline 
\multicolumn{5}{c}{\bf $\bf z_{\rm abs}$ = 0.24531} \nl
1513.86 & H I & 1215.67 & 45$\pm$9 & $1.0\pm 0.2$ \nl
1285.07 & O VI & 1031.93 & 55$\pm$6 & $5.2\pm 0.6$ \nl
1292.15 & O VI & 1037.62 & 39$\pm$6 & $7.0\pm 1.1$ \nl
\hline %\tablebreak
\multicolumn{5}{c}{\bf $\bf z_{\rm abs}$ = 0.26659} \nl
1539.75 & H I & 1215.67 & 177$\pm$12 & $4.2\pm 0.3$ \nl
1307.04 & O VI & 1031.93 & 55$\pm$8 & $5.1\pm 0.8$\nl
1314.26 & O VI & 1037.62 & 32$\pm$8 & $5.8\pm 1.3$
\enddata
\tablenotetext{a}{Observed vacuum Heliocentric wavelength of the line centroid.}
\tablenotetext{b}{Restframe vacuum wavelength from
Morton\markcite{mort91} (1991). Oscillator strengths used for these
measurements were also obtained from Morton\markcite{mort91} (1991) 
except for the Si II $\lambda$1260.42 $f-$value, which is from 
Dufton et al.\markcite{duf} (1983).}
\tablenotetext{c}{Restframe equivalent width integrated across 
all components.}
\tablenotetext{d}{Integrated apparent column density $N_{\rm a} = \int N_{\rm a}(v) dv$.}
\tablenotetext{e}{Saturated absorption line. Lower limits are 
derived from integrated apparent column densities with pixels with flux $\leq$ 
0 set to their 3$\sigma$ upper limits.}
\tablenotetext{f}{This line is probably the stronger line of the O VI  
doublet. However, the line identification is less secure than the other 
O VI lines in this table because the weaker O VI line at 1037.62 
\AA\ is blended with Milky Way S II $\lambda$1259.52 absorption. With 
higher signal-to-noise, it should be possible to confirm or refute this 
identification.}
\tablenotetext{g}{4$\sigma$ upper limit.} 
\tablenotetext{h}{A 2$\sigma$ feature is detected at the expected wavelength, 
but we do not consider this significance adequate to claim a reliable 
detection. Consequently, we set an upper limit on $N_{\rm a}$ which is the 
measured column density of the marginal line + 2$\sigma$.}
\tablenotetext{i}{The somewhat stronger C II $\lambda$1334.53 line falls in a gap between orders.}
\tablenotetext{j}{The strength and velocity extent of this line is 
surprising compared to the strength and velocity extent of the Si III 
$\lambda$1206.5 line. However, it is difficult to find convincing alternative 
identifications of this strong feature. The line may be a blend.}
\end{deluxetable}
}
%\clearpage

\end{document}